\documentclass[journal]{IEEEtran}
\usepackage{cite}

\ifCLASSINFOpdf
\else
\fi
\usepackage{graphicx}
\usepackage{bm}
\usepackage{kantlipsum,widetext}
\usepackage{multicol}
\usepackage{cite}
\usepackage{mathtools}
\usepackage{overpic}
\begin{document}
%
\title{Band structure of strained Ge$_{1-x}$Sn$_{x}$ alloy: a full-zone 30-band $k$$\cdot$$p$ model}
%
%
%

\author{Zhigang Song, Weijun Fan, Chuan Seng Tan, Qijie Wang, Donguk Nam, Dao Hua Zhang and Greg Sun
\thanks{Zhigang Song, Weijun Fan, Chuan Seng Tan, Qijie Wang, Donguk Nam and Dao Hua Zhang are with the School of Electrical and Electronic Engineering, Nanyang Technological University, Nanyang Avenue, Singapore 639798, Singapore. (email: EWJFAN@ntu.edu.sg and EDHZHANG@ntu.edu.sg).}
\thanks{Greg Sun is with the Department of Engineering, University of Massachusetts Boston, Massachusetts 02125, U.S.A}

}

\maketitle

\begin{abstract}
We extend the previous 30-band $k$$\cdot$$p$ model effectively employed for relaxed Ge$_{1-x}$Sn$_{x}$ alloy to the case of strained Ge$_{1-x}$Sn$_{x}$ alloy.
The strain-relevant parameters for the 30-band $k$$\cdot$$p$ model are obtained by using linear interpolation between the values of single crystal of Ge and Sn that are
from literatures and optimizations. We specially investigate the dependence of band-gap at $L$-valley and $\Gamma$-valley with different Sn composition under uniaxial and biaxial strain
along [100], [110] and [111] directions. The good agreement between our theoretical predictions and experimental data validates the effectiveness of our model. Our 30-band $k$$\cdot$$p$ model and relevant input parameters successfully applied to relaxed and strained Ge$_{1-x}$Sn$_{x}$ alloy offers a powerful tool for the optimization of sophisticated devices made from such alloy.

\end{abstract}

\begin{IEEEkeywords}
Ge, GeSn alloy, uniaxial strain, biaxial strain, 30-band $k$$\cdot$$p$
.
\end{IEEEkeywords}

%
\IEEEpeerreviewmaketitle

\section{Introduction}
 The Si-based optical platform has attracted considerable interests over the last decade, and its landscape is expanding rapidly with its powerful solutions such as mid-infrared lasers\cite{wirths2015lasing, millar2017mid}, infrared LEDs \cite{stange2017short} and photodetectors\cite{huang2017sn}. There is little doubt that Si photonics is becoming a mature technology as evidenced by its integration in large scale with complementary metal-oxide-semiconductor (CMOS) technology. With all the progress being made, this technology is currently being challenged, however, by the poor efficiency of light emission because of the fundamental material limitation - indirect bandgaps in Si, Ge and SiGe alloy that are employed as building materials for Si-based photonics.
 One solution that has been investigated extensively over the last decade is to alter their band structure to achieve direct band-gap through material engineering. Realizing the difference between the direct and the indirect bandgap in Si is 2.28eV while that for Ge is only 0.14eV \cite{madelung2012semiconductors,shur1996handbook}, much effort has been directed towards achieving direct bandgap by exploring strain conditions and/or material compositions in Ge or Ge-rich alloy with the goal to lower its $\Gamma$-valley below its $L$-valley.

Two approaches have been proposed and implemented in experiments. One is introduction the tensile strain in bulk Ge \cite{geiger2015uniaxially, guilloy2016germanium, sukhdeo2014direct, Boztug, Geiger,JoseR, GASSENQ201664, Saladukha} and the other is the incorporation of Sn to form Ge-rich Ge$_{1-x}$Sn${_{x}}$ alloy \cite{sun2010design10, moontragoon2012direct, wirths2016si,sun2010design1,Eckhardt, Wirths2015}. Two kinds of tensile strain have been investigated, namely biaxial and uniaxial, with the biaxial strain being introduced with the lattice-mismatched substrate \cite{JoseR} while the uniaxial strain is implemented through micro-bridges \cite{Gassenq,Donguk15, Nam55, Gupta}.
The incorporation of Sn into Ge, on the other hand, aims at obtaining direct bandgap GeSn alloy with sufficient Sn content with or without strain. In fact, significant progress has been made in the material growth of Ge$_{1-x}$Sn${_{x}}$ alloy \cite{alharthi2017study,chen2014structural,zheng2018recent} and in device fabrication \cite{wirths2015lasing, du2017investigation,millar2017mid,stange2017short,huang2017sn}. Still challenges remain with both approaches. First, the strain required to turn Ge into direct bandgap is extremely high,  $\simeq$5.6$\%$ \cite{guilloy2016germanium}, and the growth of high quality GeSn alloy with high Sn content has been proven to be difficult because of the large lattice mismatch between Ge and Sn and low solubility of Sn. The practical solution of achieving high-quality direct bandgap material could lie in combining both approaches that reduce both the tensile strain and the Sn composition required.
Even when strain is not intentionally employed, the GeSn alloy as grown below its critical thickness is subject to various degree of compressive strain either deposited on Si or on Ge substrate or buffer layer because of the lattice mismatch. Taking into account strain in GeSn alloy is absolutely essential in analysing the shift of its conduction band minima (CBM) at $L$ and $\Gamma$ valleys as well as its valence band maximum (VBM) at the $\Gamma$-valley. Needless to say an effective and efficient theoretical model is needed to guide the experimental effort in material development and device design and fabrication.

Traditional methods such as empirical pseudopotential (EPM) \cite{yahyaoui2014wave,moontragoon2012direct}, empirical tight binding (ETB) \cite{attiaoui2014indirect} and \textit{ab initio} \cite{polak2017electronic, zelazna2015electronic} have been used to study the band structure of strained Ge$_{1-x}$Sn${_{x}}$ alloy with the deformation in model-solid theory. These approaches, however, call for rather significant computational resources even for bulk materials in order to yield accurate band structures, rendering them rather inefficient to be employed for the calculation of heterostructures and/or nanostructures that are often required in Si-based photonic devices. Recently, we have developed a 30-band $k$$\cdot$$p$ model to calculate the band structure across its entire Brillouin zone (BZ) for relaxed Ge$_{1-x}$Sn${_{x}}$ alloy \cite{Zhigang}. This method not only demands far less computational resources than those traditional methods but also agrees well with experimental measurements of bandgap at $L$ and $\Gamma$-valleys. This model allows for extraction of the dependence of effective mass of electron at $L$ and $\Gamma$-valley, hole at $\Gamma$-valley, density of states around CBM and VBM, as well as Luttinger parameters on Sn composition.

In this paper, we extend our previously developed 30-band $k$$\cdot$$p$ model for relaxed Ge$_{1-x}$Sn${_{x}}$ alloy to include the strain effect. We first optimize parameters used in the 30-band $k$$\cdot$$p$ model of single-crystal strained $\alpha$-Sn as depicted in Ref. \cite{rideau2006strained} for Ge, followed by generating all input strain-relevant parameters for the 30-band $k$$\cdot$$p$ model for strained Ge$_{1-x}$Sn${_{x}}$ alloy by a linear interpolation. The resulting Hamiltonian of the strained Ge$_{1-x}$Sn${_{x}}$ alloy is thus obtained by combining the Hamiltonian of strained with that of relaxed. Based on this model, we study the band gap variation of Ge$_{1-x}$Sn${_{x}}$ alloy under biaxial and uniaxial strain along [100], [110] and [111] directions for different Sn compositions. Finally, we compare our theoretical predictions with available experimental data in the literatures. The good agreement suggests that the 30-band $k$$\cdot$$p$ model can serve as an accurate and efficient design tool for the optimization of sophisticated devices made from either relaxed or strained Ge$_{1-x}$Sn${_{x}}$ alloys in hetero- and nanostructures.

\section{Model}

Following the strain formalism used in the 8-band $k$$\cdot$$p$ model, we extend it to the case of 30-band $k$$\cdot$$p$ Hamiltonian with the perturbation induced by the strain written as \cite{rideau2006strained}

\begin{widetext}
\begin{equation}
W_{kp}^{30}=\left(
\begin{array}{cccccccc}
W_{\Gamma _{2^{\prime u}}}^{2\times 2} & P_{4}W_{k}^{2\times 6} & W_{\Gamma
_{2^{\prime u}}\Gamma _{12^{\prime }}}^{2\times 4} & 0 & 0 & W_{\Gamma
_{2^{\prime u}}\Gamma _{15}}^{2\times 6} & W_{\Gamma _{2^{\prime u}}\Gamma
_{2^{\prime l}}}^{2\times 2} & P_{3}W_{k}^{2\times 6} \\
& W_{\Gamma _{25^{\prime u}}}^{6\times 6} & R_{2}W_{k}^{6\times 4} &
W_{\Gamma _{25^{\prime u}}\Gamma _{1^{u}}}^{6\times 2} & W_{\Gamma
_{25^{\prime u}}\Gamma _{1^{l}}}^{6\times 2} & Q_{2}W_{k}^{6\times 6} &
P_{2}W_{k}^{6\times 2} & W_{\Gamma _{25^{\prime l}}\Gamma _{25^{\prime
u}}}^{6\times 6} \\
&  & W_{\Gamma _{12^{\prime }}}^{4\times 4} & 0 & 0 & W_{\Gamma _{12^{\prime
}}\Gamma _{15}}^{4\times 6} & W_{\Gamma _{12^{\prime }}\Gamma _{2^{\prime
l}}}^{4\times 2} & R_{1}W_{k}^{4\times 6} \\
&  &  & W_{\Gamma _{1^{u}}}^{2\times 2} & W_{\Gamma _{1^{u}}\Gamma
_{1^{l}}}^{2\times 2} & T_{1}W_{k}^{2\times 6} & 0 & W_{\Gamma
_{1^{u}}\Gamma _{25^{\prime l}}}^{2\times 6} \\
&  &  &  & W_{\Gamma _{1^{l}}}^{2\times 2} & T_{2}W_{k}^{2\times 6} & 0 &
W_{\Gamma _{1^{l}}\Gamma _{25^{\prime l}}}^{2\times 6} \\
&  &  &  &  & W_{\Gamma _{15}}^{6\times 6} & W_{\Gamma _{15}\Gamma
_{2^{\prime l}}}^{6\times 2} & Q_{1}W_{k}^{6\times 6} \\
&  &  &  &  &  & W_{\Gamma _{2^{\prime l}}}^{2\times 2} &
P_{1}W_{k}^{2\times 6} \\
&  &  &  &  &  &  & W_{\Gamma _{25^{\prime l}}}^{6\times 6}%
\end{array}%
\right)
\end{equation}

where two kinds of terms, $k$-dependent and $k$-independent, can be distinguished and appear as follows. The $k$-independent $W_{\Gamma}$ are written as:
\begin{equation}
W_{\Gamma }^{6\times 6}=\left(
\begin{array}{cc}
W_{\Gamma }^{3\times 3} & 0 \\
0 & W_{\Gamma }^{3\times 3}%
\end{array}%
\right)
\end{equation}

\begin{equation}
  W_{\Gamma }^{3\times 3}=\left(
\begin{array}{ccc}
l\epsilon _{xx}+m\left( \epsilon _{yy}+\epsilon _{zz}\right)  & n\epsilon
_{xy} & n\epsilon _{xz} \\
n\epsilon _{xy} & l\epsilon _{yy}+m\left( \epsilon _{xx}+\epsilon
_{zz}\right)  & n\epsilon _{yz} \\
n\epsilon _{xz} & n\epsilon _{yz} & l\epsilon _{zz}+m\left( \epsilon
_{xx}+\epsilon _{yy}\right)
\end{array}%
\right)
\end{equation}

\begin{equation}
W_{\Gamma }^{4\times 2}=g_{\Gamma }\left(
\begin{array}{cc}
\sqrt{3}\left( \epsilon _{yy}-\epsilon _{zz}\right) & 0 \\
2\epsilon _{xx}-\epsilon _{yy}-\epsilon _{zz} & 0 \\
0 & \sqrt{3}\left( \epsilon _{yy}-\epsilon _{zz}\right) \\
0 & 2\epsilon _{xx}-\epsilon _{yy}-\epsilon _{zz}%
\end{array}%
\right)
\end{equation}

\begin{equation}
W_{\Gamma }^{2\times 2}=a_{\Gamma }\left(
\begin{array}{cc}
\epsilon _{xx}+\epsilon _{yy}+\epsilon _{zz} & 0 \\
0 & \epsilon _{xx}+\epsilon _{yy}+\epsilon _{zz}%
\end{array}%
\right)
\end{equation}

\begin{equation}
W_{\Gamma }^{2\times 6}=f_{\Gamma }\left(
\begin{array}{cccccc}
\epsilon _{yz} & \epsilon _{xz} & \epsilon _{xy} & 0 & 0 & 0 \\
0 & 0 & 0 & \epsilon _{yz} & \epsilon _{xz} & \epsilon _{xy}%
\end{array}%
\right)
\end{equation}

\begin{equation}
W_{\Gamma }^{4\times 6}=h_{\Gamma }\left(
\begin{array}{cccccc}
0 & \sqrt{3}\epsilon _{xz} & -\sqrt{3}\epsilon _{xy} & 0 & 0 & 0 \\
2\epsilon _{yz} & -\epsilon _{xz} & -\epsilon _{xy} & 0 & 0 & 0 \\
0 & 0 & 0 & 0 & \sqrt{3}\epsilon _{xz} & -\sqrt{3}\epsilon _{xy} \\
0 & 0 & 0 & 2\epsilon _{yz} & -\epsilon _{xz} & -\epsilon _{xy}%
\end{array}%
\right)
\end{equation}

\begin{equation}
W_{\Gamma _{12}}^{4\times 4}=\left(
\begin{array}{cccc}
A\epsilon _{xx}+B\left( \epsilon _{yy}+\epsilon _{zz}\right) & E\left(
\epsilon _{yy}-\epsilon _{zz}\right) & 0 & 0 \\
E\left( \epsilon _{yy}-\epsilon _{zz}\right) & A\epsilon _{xx}+B\left(
\epsilon _{yy}+\epsilon _{zz}\right) & 0 & 0 \\
0 & 0 & A\epsilon _{xx}+B\left( \epsilon _{yy}+\epsilon _{zz}\right) &
E\left( \epsilon _{yy}-\epsilon _{zz}\right) \\
0 & 0 & E\left( \epsilon _{yy}-\epsilon _{zz}\right) & A\epsilon
_{xx}+B\left( \epsilon _{yy}+\epsilon _{zz}\right)%
\end{array}%
\right)
\end{equation}

where the coefficients
\begin{eqnarray*}
A &=&6\left( b_{12}-d_{12}\right) \\
B &=&3\left( a_{12}+b_{12}-2c_{12}\right) \\
C &=&2\left( 2a_{12}-4c_{12}+b_{12}+d_{12}\right) \\
D &=&5b_{12}-2c_{12}-4d_{12}+a_{12} \\
E &=&\sqrt{3}\left( 2c_{12}-2d_{12}-a_{12}+b_{12}\right)
\end{eqnarray*}

The $k$-independent $W_{k}$ can be written as:
\begin{equation}
W_{k}^{6\times 6}=-\sum_{i}\left(
\begin{array}{cccccc}
0 & \epsilon _{zi}k_{i} & \epsilon _{yi}k_{i} & 0 & 0 & 0 \\
\epsilon _{iz}k_{i} & 0 & \epsilon _{xi}k_{i} & 0 & 0 & 0 \\
\epsilon _{iy}k_{i} & \epsilon _{ix}k_{i} & 0 & 0 & 0 & 0 \\
0 & 0 & 0 & 0 & \epsilon _{zi}k_{i} & \epsilon _{yi}k_{i} \\
0 & 0 & 0 & \epsilon _{iz}k_{i} & 0 & \epsilon _{xi}k_{i} \\
0 & 0 & 0 & \epsilon _{iy}k_{i} & \epsilon _{ix}k_{i} & 0%
\end{array}%
\right)
\end{equation}

\begin{equation}
  W_{k}^{4\times 6}=-\sum_{i}\left(
\begin{array}{cccccc}
0 & \sqrt{3}\epsilon _{iy}k_{i} & -\sqrt{3}\epsilon _{iz}k_{i} & 0 & 0 & 0
\\
2\epsilon _{ix}k_{i} & -\epsilon _{iy}k_{i} & -\epsilon _{iz}k_{i} & 0 & 0 &
0 \\
0 & 0 & 0 & 0 & \sqrt{3}\epsilon _{iy}k_{i} & -\sqrt{3}\epsilon _{iz}k_{i}
\\
0 & 0 & 0 & 2\epsilon _{ix}k_{i} & -\epsilon _{iy}k_{i} & -\epsilon
_{iz}k_{i}%
\end{array}%
\right)
\end{equation}

\begin{equation}
W_{k}^{2\times 6}=-\sum_{i}\left(
\begin{array}{cccccc}
\epsilon _{ix}k_{i} & \epsilon _{iy}k_{i} & \epsilon _{iz}k_{i} & 0 & 0 & 0
\\
0 & 0 & 0 & \epsilon _{ix}k_{i} & \epsilon _{iy}k_{i} & \epsilon _{iz}k_{i}%
\end{array}%
\right)
\end{equation}

\end{widetext}

Since the strain tensor is directly influenced by the kind of strain and corresponding direction, we have considered three different directions [100], [110] and [111] for biaxial and uniaxial strain.

Suppose the Ge$_{1-x}$Sn$_{x}$ alloy is grown on [100], [110] and [111] substrates and their corresponding lattice constants are $a_{s}$ and $a_{x}$, the in-plane strain $\varepsilon_{||}=a_{s}/a_{x}-1$ and the vertical strain $\varepsilon_{\bot}=-2C_{12}/C_{11}\varepsilon_{||}$. The biaxial strain tensor of the three directions is therefore:
\begin{equation}
\epsilon _{\lbrack 100]}=\left(
\begin{array}{ccc}
\varepsilon _{\bot } & 0 & 0 \\
0 & \varepsilon _{||} & 0 \\
0 & 0 & \varepsilon _{||}%
\end{array}%
\right)
\end{equation}

\begin{equation}
\epsilon _{\lbrack 110]}=\left\{
\begin{array}{c}
\varepsilon _{xx}=\varepsilon _{yy}=\left( \varepsilon _{\bot }+\varepsilon
_{||}\right) /2 \\
\varepsilon _{zz}=\varepsilon _{||} \\
\varepsilon _{xy}=\left( \varepsilon _{\bot }-\varepsilon _{||}\right) /2\\
\varepsilon _{yz}=\varepsilon _{xz}=0 %
\end{array}%
\right.
\end{equation}

\begin{equation}
\epsilon _{\lbrack 111]}=\left\{
\begin{array}{c}
\varepsilon _{xx}=\varepsilon _{yy}=\varepsilon _{zz}=\left( \varepsilon
_{\bot }+2\varepsilon _{||}\right) /3 \\
\varepsilon _{xy}=\varepsilon _{yz}=\varepsilon _{xz}=\left( \varepsilon
_{\bot }-\varepsilon _{||}\right) /3%
\end{array}%
\right.
\end{equation}
Similarly, the uniaxial strain tensor can be written as:
\begin{equation}
\varepsilon _{\lbrack 100]}=P\left(
\begin{array}{ccc}
s_{11} & 0 & 0 \\
0 & s_{12} & 0 \\
0 & 0 & s_{12}%
\end{array}%
\right)
\end{equation}

\begin{equation}
\varepsilon _{\lbrack 110]}=\frac{P}{2}\left(
\begin{array}{ccc}
s_{11}+s_{12} & s_{44}/2 & 0 \\
s_{44}/2 & s_{11}+s_{12} & 0 \\
0 & 0 & 2s_{12}%
\end{array}%
\right)
\end{equation}

\begin{equation}
\varepsilon _{\lbrack 111]}=\frac{P}{3}\left(
\begin{array}{ccc}
s_{11}+2s_{12} & s_{44}/2 & s_{44}/2 \\
s_{44}/2 & s_{11}+2s_{12} & s_{44}/2 \\
s_{44}/2 & s_{44}/2 & s_{11}+2s_{12}%
\end{array}%
\right)
\end{equation}
where $P$ is the uniaxial stress and $s_{ij}$$'s$ are elastic compliances and can be calculated as:
\begin{subequations}
\begin{align}
s_{11} &=\frac{C_{11}+C_{12}}{(C_{11}-C_{12})(C_{11}+2C_{12})} \\
s_{12} &=\frac{-C_{12}}{(C_{11}-C_{12})(C_{11}+2C_{12})} \\
s_{44} &=\frac{1}{C_{44}}
\end{align}
\end{subequations}

All the elastic constants $C_{11}$, $C_{12}$ and $C_{44}$ of Ge$_{1-x}$Sn${_{x}}$ alloy are calculated by linear
interpolation between values for single crystal Ge and Sn found in \cite{madelung2012semiconductors,shur1996handbook}.

\begin{table*}
\begin{center}
\caption{\label{tab:table4}Strain perturbation matrix coefficients expressed in eV of Ge$_{1-x}$Sn$_{x}$. }
\begin{tabular}{cccccccc}
\hline
\hline
Symbols                 & Ge$_{1-x}$Sn$_{x}$  &Symbols               &Ge$_{1-x}$Sn$_{x}$ &Symbols                               & Ge$_{1-x}$Sn$_{x}$  &Symbols                      &Ge$_{1-x}$Sn$_{x}$\\ \hline
$l_{\Gamma_{25^{l}}}$   & -3.8+7.276$x$       &$a_{12}$              &6.815-6.358$x$     &$l_{\Gamma_{25^{l}},\Gamma_{25^{u}}}$ &-24.139-18.805$x$    &$a_{\Gamma_{2^{l}},\Gamma_{2^{u}}}$ &-1.211 -1.381$x$ \\
$m_{\Gamma_{25}^{l}}$   & 4.9-4.947$x$        &$b_{12}$              &6.798-7.676$x$     &$m_{\Gamma_{25^{l}},\Gamma_{25^{u}}}$ &-0.124+0.513$x$      &$a_{\Gamma_{1^{l}},\Gamma_{1^{u}}}$ &-5.927+14.470$x$\\
$n_{\Gamma_{25}^{l}}$   & -9.527+12.684$x$    &$c_{12}$              &7.745-7.553$x$     &$n_{\Gamma_{25^{l}},\Gamma_{25^{u}}}$ &-0.112-2.257$x$      &$g_{\Gamma_{12},\Gamma_{2^{u}}}$    &-5.000+3.279$x$ \\
$l'_{\Gamma_{15}}$      & 6.026+34.467$x$     &$d_{12}$              &4.858-6.222$x$     &$f_{\Gamma_{1^{u}},\Gamma_{25^{u}}}$  &11.220-5.050$x$      &$g_{\Gamma_{12},\Gamma_{2^{l}}}$    &-5.354-4.225$x$ \\
$m'_{\Gamma_{15}}$      & 0.762-40.103$x$     &$a_{\Gamma_{2^{l}}}$  &-7.181+4.152$x$    &$f_{\Gamma_{1^{l}},\Gamma_{25^{l}}}$  &-7.666-3.630$x$   \\
$n'_{\Gamma_{15}}$      &-10.134+8.697$x$     &$a_{\Gamma_{2^{u}}}$  &4.490+11.955$x$    &$f_{\Gamma_{1^{u}},\Gamma_{25^{l}}}$  &-12.210+19.503$x$ \\
$l''_{\Gamma_{25^{u}}}$ & -20.692+14.536$x$   &$a_{\Gamma_{1^{l}}}$  &14.171-15.427$x$   &$f_{\Gamma_{15},\Gamma_{2^{l}}}$      &-22.242+38.110$x$ \\
$m''_{\Gamma_{25^{u}}}$ & 9.119-8.853$x$      &$a_{\Gamma_{1^{u}}}$  &-0.492-16.491$x$   &$f_{\Gamma_{15},\Gamma_{2^{u}}}$      &19.925+24.338$x$  \\
$n''_{\Gamma_{25^{u}}}$ & 0.481+0.054$x$ \\

\hline
\hline
\end{tabular}
\end{center}
\end{table*}

\begin{figure}
\includegraphics[width=3.4in,height=5.2in]{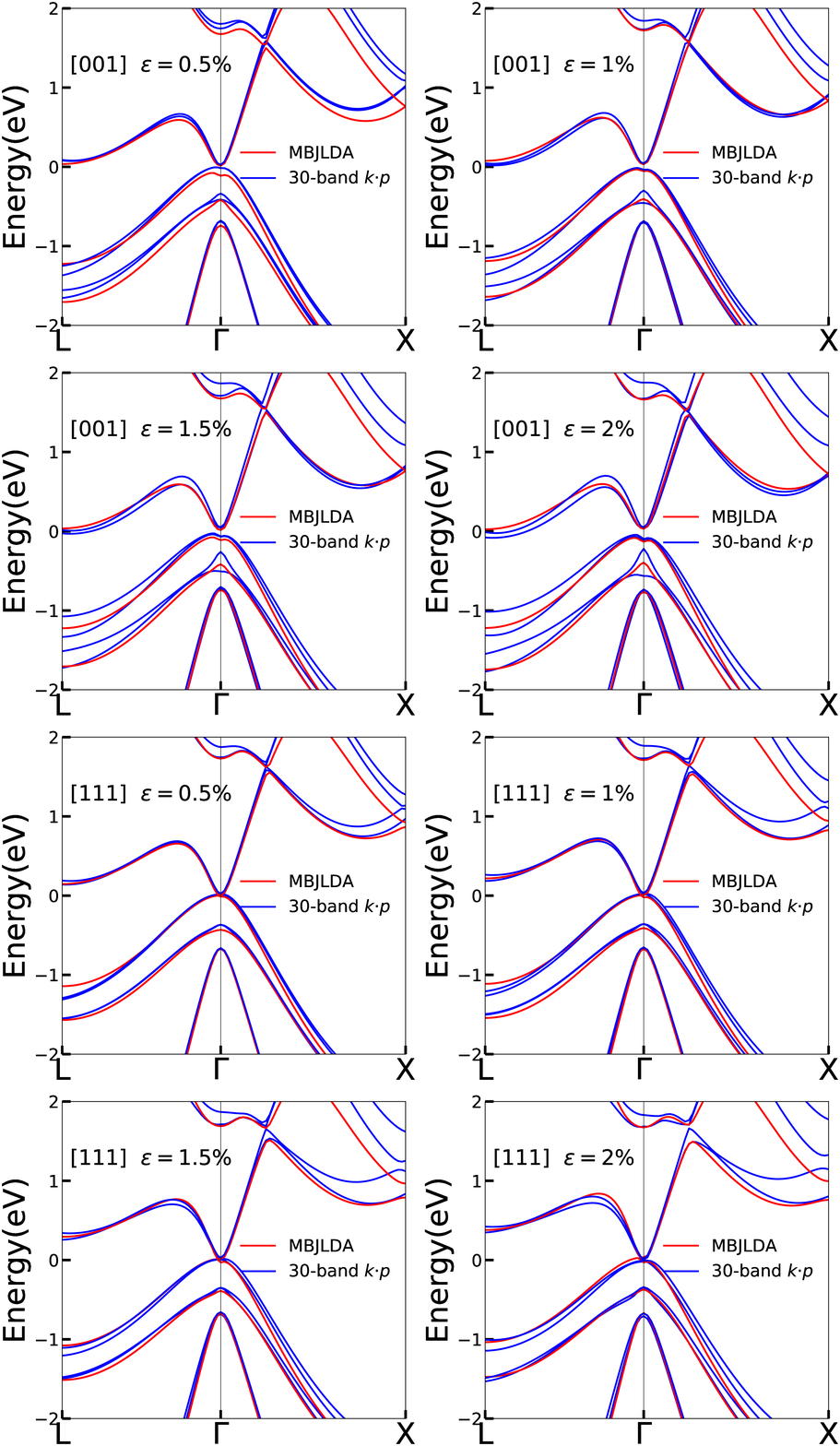}
\caption{Sn electronic band structure obtained from the MBJLDA method (red) and fitted by the 30-band $k$$\cdot$$p$ model (blue) at room temperature under different biaxial tensile strain along [001] and [111].}
\label{fig1}
\end{figure}

\section{Results and Discussion}
Since the input parameters of the 30-band model for strained Ge have been previously optimized \cite{rideau2006strained},
we only need to optimize the input parameters for Sn. For convenience,
we assume that the VBM is at potential zero in the absence of strain and that
all other values are referenced to it in all calculations of energy.

With the help of $ab$ $initio$ method and hill climbing technique \cite{Zhigang} that are used in the 30-band parameters optimization process, the band structures of Sn in the presence of biaxial tensile strains along [001] and [111] are calculated. In order to avoid underestimating the band gap, a hybrid functional based on modified Becke-Johnson local density approximation (MBJLDA) \cite{becke2006simple} is implemented. We can see from Fig. \ref{fig1} that the two band structures  obtained by MBJLDA and the 30-band model are nicely matched globally
across the full BZ except around the $X$ point which is not our interests for the Ge$_{1-x}$Sn$_{x}$ alloy. The good overall agreement between
them clearly validates the effectiveness of the 30-band $k$$\cdot$$p$ in the presence of strain. Thus, we can now derive the input parameters of strained Ge$_{1-x}$Sn${_{x}}$ alloy by linear interpolation between Ge and Sn. All the strained input parameters are listed in Table \ref{tab:table4}.
\begin{figure*}
\includegraphics[width=7.0in,height=2in]{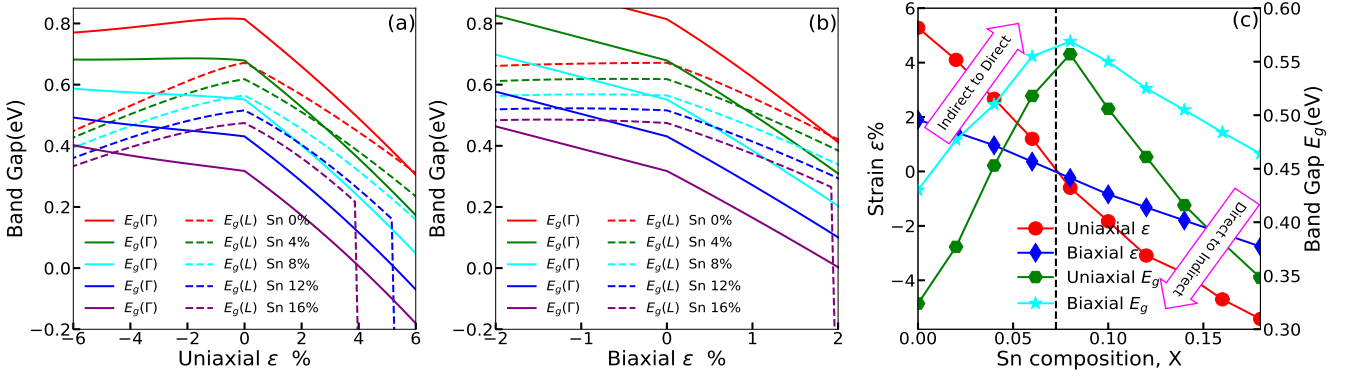}
\caption{The bandgaps at $\Gamma$-valley (solid) and $L$-valley (dashed) of Ge$_{1-x}$Sn$_{x}$ alloy with different Sn compositions vs uniaxial strain (a) and biaxial strain (b) along [100] direction. Ge$_{1-x}$Sn$_{x}$ alloy from indirect-to-direct bandgap for Sn composition below that of the crossover point for relaxed Ge$_{1-x}$Sn$_{x}$ marked at the vertical line and the compressive strain required for Sn composition above the crossover point to transition from direct-to-indirect. The green and light blue curves show the bandgap of Ge$_{1-x}$Sn$_{x}$ alloy at the Sn composition in the presence of the corresponding uniaxial and biaxial strain, respectively. }
\label{fig2}
\end{figure*}
Compared with the relaxed case, the strain may or may not lift the degeneracy of the four $L$-valleys depending on the direction it is applied. Similar phenomena around $X$-valley have been observed in Si \cite{rideau2006strained}. In order to distinguish them, we shall label the four $L$-valleys as $L_{1}$ ([111]), $L_{2}$ ([11$\overline{1}$]) $L_{3}$ ([1$\overline{1}$1]) and $L_{4}$ ([$\overline{1}$11]). We shall now present the results of the band-gap at the $L$ and $\Gamma$-valleys under strain along [100], [110] and [111] directions.

\begin{figure*}
\includegraphics[width=6.8in,height=2in]{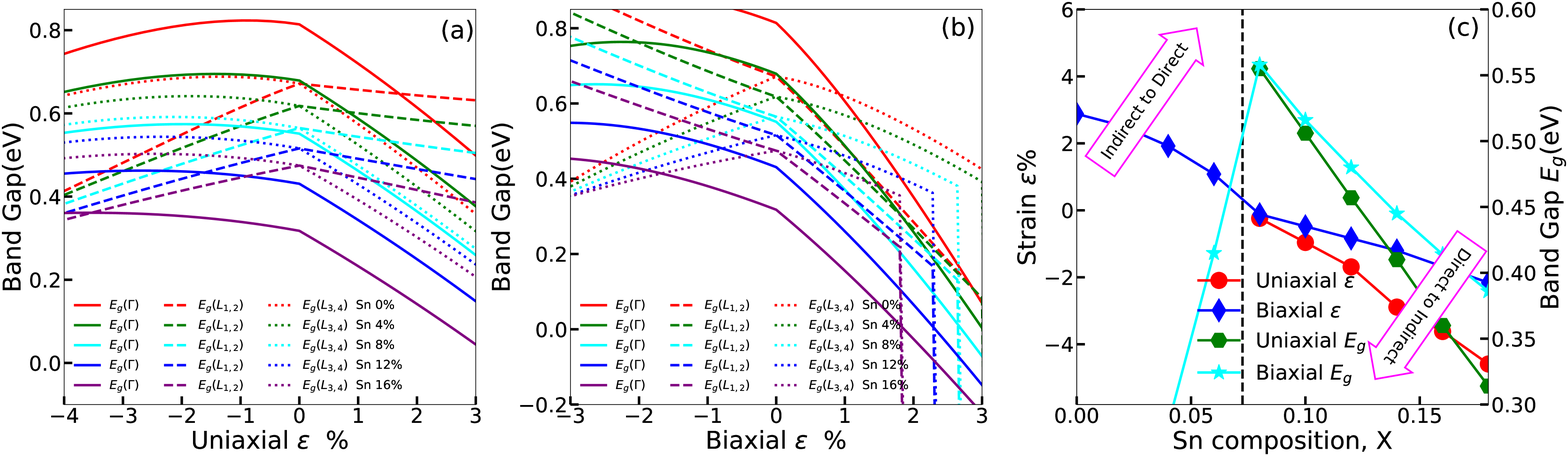}
\caption{The bandgaps at $\Gamma$-valley (solid) and $L$-valley (dashed and dotted) of Ge$_{1-x}$Sn$_{x}$ alloy with different Sn compositions vs uniaxial strain (a) and biaxial strain (b) along [110] direction. Different from the [100] case, [110] strain can lift the degeneracy of the $L$-valleys into 2 groups, $g_{1}^{110}$ (dashed) and $g_{2}^{110}$ (dotted), each is of 2-degeneracy. (c) The [110] tensile strain of uniaxial (red) and biaxial (blue) required to turn the Ge$_{1-x}$Sn$_{x}$ alloy from indirect-to-direct bandgap for Sn composition below that of the crossover point for relaxed Ge$_{1-x}$Sn$_{x}$ marked at the vertical line and the compressive strain required for Sn composition above the crossover point to transition from direct-to-indirect. The green and light blue curves show the bandgap of Ge$_{1-x}$Sn$_{x}$ alloy at the Sn composition in the presence of the corresponding uniaxial and biaxial strain, respectively.
}
\label{fig3}
\end{figure*}

\begin{figure*}
\includegraphics[width=6.8in,height=2in]{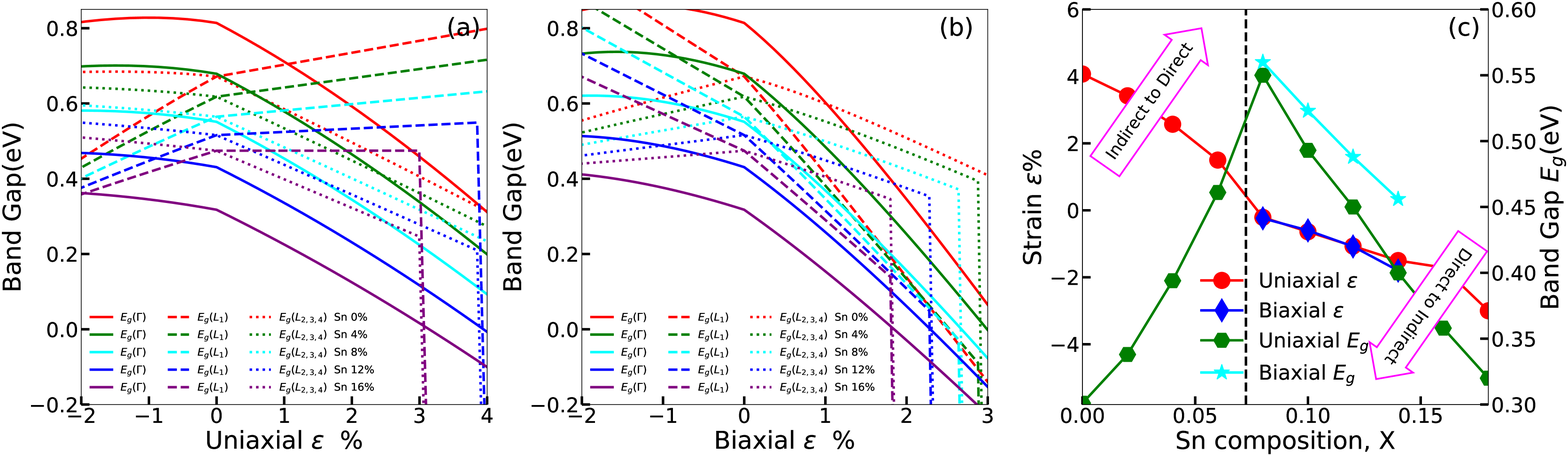}
\caption{The bandgaps at $\Gamma$-valley (solid) and $L$-valley (dashed and dotted) of Ge$_{1-x}$Sn$_{x}$ alloy with different Sn compositions vs uniaxial strain (a) and biaxial strain (b) along [111] direction. Here the degeneracy of the $L$-valley into 1-degeneracy at $L_{1}$-valley (dashed) and 3-degeneracy at $L_{2}$, $L_{3}$ and $L_{4}$ (dotted). (c) The [111] tensile strain of uniaxial (red) and biaxial (blue) required to turn the Ge$_{1-x}$Sn$_{x}$ alloy from indirect-to-direct bandgap for Sn composition below that of the crossover point for relaxed Ge$_{1-x}$Sn$_{x}$ marked at the vertical line and the compressive strain required for Sn composition above the crossover point to transition from direct-to-indirect. The green and light blue curves show the bandgap of Ge$_{1-x}$Sn$_{x}$ alloy at the Sn composition in the presence of the corresponding uniaxial and biaxial strain, respectively.}
\label{fig4}
\end{figure*}

The bandgaps at $\Gamma$ and $L$-valley as a function of the uniaxial and biaxial strain along [100] direction for a range of Sn compositions are shown in Fig. \ref{fig2}(a) and (b), respectively.
For both types of the strain along [100], the degeneracy at $L$-valley is not lifted. It can be seen in Fig. \ref{fig2}(a) that under both compressive and tensile uniaxial strain, the bandgap at $L$-valley decreases with the increase of strain of either type, but the bandgap at $\Gamma$-valley exhibits monotonic behaviour throughout the range of uniaxial strain from compressive to tensile. For the biaxial strain, however, the bandgap at $L$-valley hardly changes under the influence of the compressive strain as shown in Fig. \ref{fig2}(b). For both uniaxial and biaxial tensile strain, the decreasing rate of the $L$-valley band-gap is indeed slower than that of the $\Gamma$-valley as shown in Fig. \ref{fig2}(a) and (b). Therefore, the CBM at $\Gamma$-valley can be lower than that at $L$-valley when the strength of tensile strain exceeds a critical value for the Ge$_{1-x}$Sn$_{x}$ alloy, transitioning Ge$_{1-x}$Sn$_{x}$ into a direct bandgap material. At a fixed Sn composition, tensile strain can also push the CBM at $\Gamma$-valley across with the VBM, causing the tensile-strained Ge$_{1-x}$Sn$_{x}$ alloy to have zero bandgap at $\Gamma$-valley and beyond such that the CBM resides below the VBM at $\Gamma$-valley. When this occurs, the conduction and valence band reverse in energy and it becomes meaningless to continue referring to the bandgap in $L$-valley (thus the vertical dashed lines in Fig. \ref{fig2}(a) and (b)).

It is not difficult to see that the Sn composition required for the Ge$_{1-x}$Sn$_{x}$ alloy to become direct bandgap under the tensile strain is less than when it is unstrained, as marked by the dashed vertical line in Fig. 2(c), to the left side of which, the red and blue curves give the tensile strain of either uniaxial or biaxial required at the Sn composition to turn Ge$_{1-x}$Sn$_{x}$ alloy from indirect to direct bandgap, respectively. On the right side of the vertical line where the Sn composition is beyond what it takes for relaxed Ge$_{1-x}$Sn$_{x}$ alloy to be direct bandgap, the two curves (red and blue) actually indicate the amount of compressive strain that will turn the material from direct back to indirect bandgap. The green and light blue curves show the crossover bandgap (bandgap at$\Gamma$ and $L$-valley being equal) of Ge$_{1-x}$Sn$_{x}$ alloy at the Sn composition in the presence of the corresponding uniaxial and biaxial strain, respectively. This phenomenon not only take place [100] strain, it can also be found in [110] and [111] directions as shown in Figs. 3 and 4, respectively.

When strain along [110] is introduced, the degeneracy of L-valleys is lifted and the four L-valleys can be classified into two groups: $L_{1}$ and $L_{2}$ belong to group $g_{1}^{110}$ while $L_{3}$ and $L_{4}$ to group $g_{2}^{110}$. We use dashed and dotted curves to represent the two groups. Unlike the situation in which strain is applied in the [100] direction as shown in Fig. \ref{fig2}, the two groups exhibit rather their own distinctive behavior different dependence on uniaxial and biaxial strain. The band gap at $L_{1,2}$ is nearly unchanged while that of $L_{3,4}$ decreases rapidly under uniaxial tensile strain as shown in Fig. \ref{fig3}(a). The opposite is true for the biaxial tensile strain under which the change of $L_{3,4}$-group is slower than that of $L_{1,2}$-group shown in Fig. \ref{fig3}(b). Once again the vertical lines for $L_{1,2}$-group in Fig. \ref{fig3}(a) and (b) indicate the reversal of energy between CBM and VBM at $\Gamma$-valley as depicted in Ref. \cite{ESCALANTE2018223} for single crystal Ge. Once again, it can be established that the Sn composition required for the Ge$_{1-x}$Sn$_{x}$ alloy to make the indirect-to-direct transition is reduced when tensile strain of either type is introduced along [110] as shown in Fig. \ref{fig3}(c) and the compressive strain can also turn the direct bandgap Ge$_{1-x}$Sn$_{x}$ alloy when it is relaxed to indirect. The green and light blue curves show the crossover
bandgap (bandgap at $\Gamma$ and $L$-valley being equal) at the Sn composition in the presence of the
corresponding uniaxial and biaxial strain, respectively.

For the [111] strain, the four L-valleys once again can be classified into two groups: $L_{1}$ belongs to group $g^{111}_{1}$ and $L_{2}$ $L_{3}$ and $L_{4}$ belong to group $g^{111}_{2}$. Similar to the situation of [110] strain, their dependence on uniaxial and biaxial strain are opposite of each other as shown in Fig. \ref{fig4}(a) and (b). The amount of Sn composition required for indirect-to-direct crossover is reduced by the application of tensile strain along [111] as shown in Fig. \ref{fig4}(c) (left side of the vertical line) and, once again, compressive strain can make relaxed direct bandgap Ge$_{1-x}$Sn$_{x}$ alloy into indirect (right side of the vertical line). The green and light blue curves show the crossover
bandgap (bandgap at $\Gamma$ and $L$-valley being equal) at the Sn composition in the presence of the
corresponding uniaxial and biaxial strain, respectively.where

These different behaviours of the four degenerate $L$-valley groups can be explained by the deformation in the model-solid theory \cite{Pollak_strain, VandeWalle}. According to that theory, the strain along [110] leads to $\Delta E_{L_{1,2}}=\frac{2}{3}\Xi_{u}^{L}\varepsilon_{xy}$ and $\Delta E_{L_{3,4}}=-\frac{2}{3}\Xi_{u}^{L}\varepsilon_{xy}$, while the strain along [111] leads to $\Delta E_{L_{1}}=2\Xi_{u}^{L}\varepsilon_{xy}$ and $\Delta E_{L_{2,3,4}}=-\frac{2}{3}\Xi_{u}^{L}\varepsilon_{xy}$, respectively. When uniaxial tensile strain is applied along [110] or [111], P is positive which means $\varepsilon_{xy}$ is also positive. Under biaxial tensile strain along [110] or [111], however, $\varepsilon_{xy}$ is negative. Therefore, the strain-induced variation of the two groups for [110] and [111] are always opposite of one another. It should therefore be pointed out that special attention needs to be paid when using strain along [110] or [111] direction on Ge$_{1-x}$Sn$_{x}$ alloy before it becomes direct bandgap because the lift of the degeneracy increases the complexity of the band structure for electron. Clearly for simplicity in strain engineering, tensile strain along [100] direction is the best choice. It should also be pointed out that Ge$_{1-x}$Sn$_{x}$ alloys grown on either Si or Ge substrates are always subject to compressive strain, negatively impeding the transition of this alloy to direct bandgap material. Other options in substrates or buffer layers with larger lattice constants should to be considered, including in order to introduce tensile strain in Ge$_{1-x}$Sn$_{x}$ alloy, e.g. relaxed Ge$_{1-y}$Sn$_{y}$ alloy buffer layers of higher Sn compositions ($y>x$).

\begin{figure}
\includegraphics[width=3.4in]{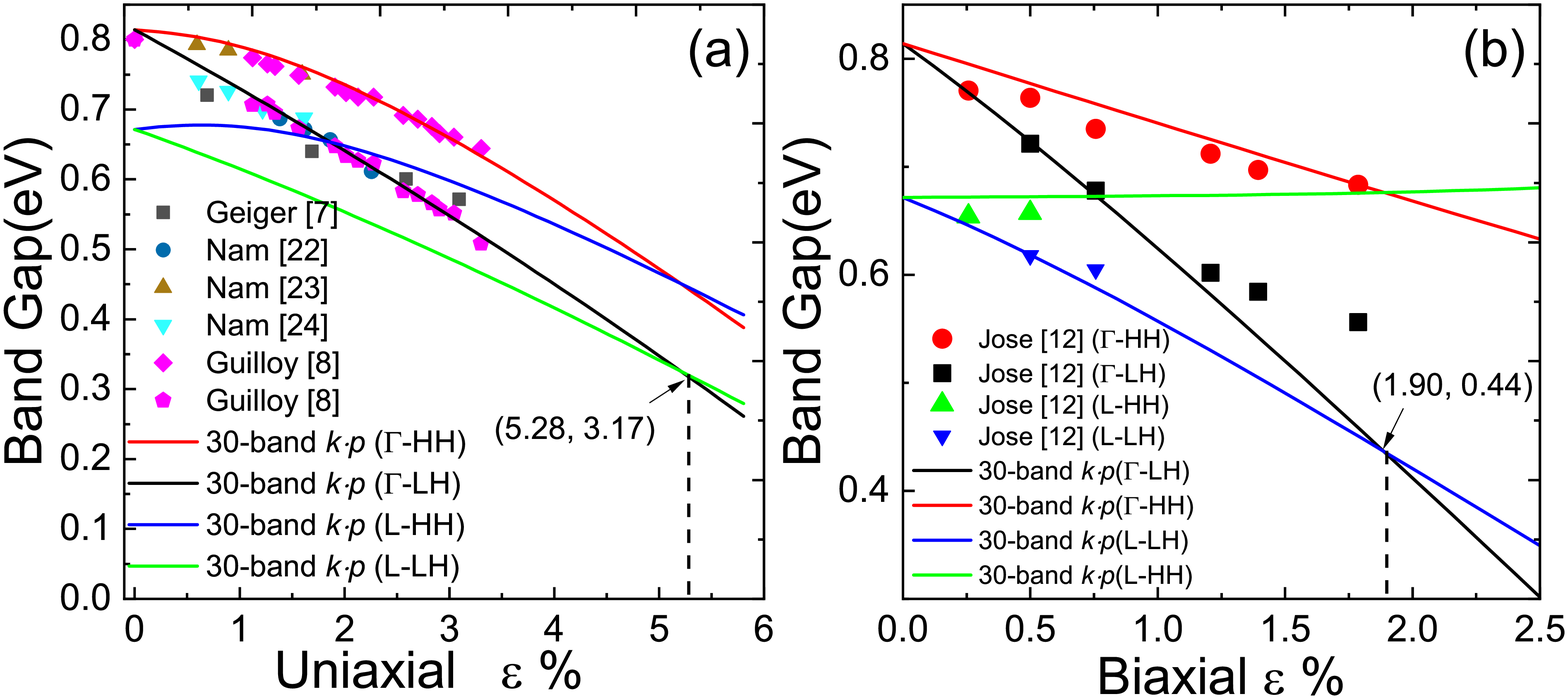}
\caption{Comparison of bandgap prediction calculated by 30-band $k$$\cdot$$p$ model with the published experimental data under (a) uniaxial (b) biaxial tensile strain.}
\label{fig5}
\end{figure}

To examine the validity of our 30-band $k$$\cdot$$p$ model that in accounting for the strain effect, let us now compare the results of theoretical prediction from our 30-band $k$$\cdot$$p$ model with the existing experimental data on single crystal Ge in the literature. Comparison of our calculation results (solid lines) with the experimental data (scattered dots) for uniaxial and biaxial strain is shown in Fig. 5(a) and 5(b), respectively. The agreement is rather nearly perfect remarkable for uniaxial results. For instance, the indirect-to-direct crossover point for Ge determined from our model for the uniaxial tensile strain is 5.28$\%$ and the corresponding bandgap is 0.31eV, results extrapolated from experimental data in \cite{guilloy2016germanium} are 5.6$\%$ and 0.305eV, respectively. The experimental data on Ge under biaxial strain are not as abundant in the literature. For what was available, the agreement is reasonable as shown in Fig. \ref{fig5}(b).


It should be noted  that our labels of heavy hole (HH) and light hole (LH) are opposite of Ref. \cite{guilloy2016germanium} and labels of HH and LH are inconsistent in Ref. \cite{wada2015photonics, ESCALANTE2018223, Donguk15, Nam55}. The inconsistency in labelling the HH and LH bands is often times the result of employment of incorrect model in interpreting the experimental data.

\begin{figure}
\includegraphics[width=3.4in]{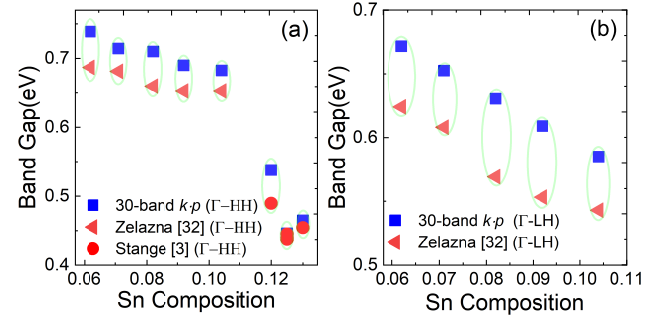}
\caption{Bandgaps relative to (a) HH band and (b) LH band calculated by the 30-band model compared (blue squares) with the experimental data (red triangles \cite{zelazna2015electronic} and red circles \cite{stange2017short }) under biaxial compressive strain.}
\label{fig6}
\end{figure}

Finally, we compare our 30-band model with experimental measurements on the bandgap of Ge$_{1-x}$Sn$_{x}$ alloy samples of different Sn compositions under various degrees of strain. Using the published Sn composition of the sample as well as the amount of strain under which the measurements are conducted \cite{stange2017short,zelazna2015electronic} as the input parameters, the predictions of bandgap calculated with the 30-band model are shown along with the experimental results measured between CBM at $\Gamma$-valley and HH and LH band in Fig. \ref{fig6} (a) and (b), respectively. Once again the agreement  in reasonable with the experimental results for different Sn compositions under various degrees of strain albeit our prediction of the bandgap is consistently higher the published results. As more experimental data become available in the literature, the 30-band model taking into account of strain can be further refined by fine tuning its input parameters and its accuracy is expected to improve.

\section{Conclusion}

In summary, we have extended our previous 30-band $k$$\cdot$$p$ model on relaxed Ge$_{1-x}$Sn$_{x}$ to include the strain effect on the band structures. Based on this strained model, the variation of Ge$_{1-x}$Sn$_{x}$ alloy band gap at $L$-valley and $\Gamma$-valley under uniaxial strain and biaxial strain along [100] [110] [111] are systematically studied. We have compared the results obtained from the 30-band $k$$\cdot$$p$ model with the published experimental data under various conditions of strain. The good agreement suggests the 30-band $k$$\cdot$$p$ model is an effective method in calculating band structure of Ge$_{1-x}$Sn$_{x}$ in the presence of strain and can serve as a powerful tool in the design of complex photonic devices made from the Ge$_{1-x}$Sn$_{x}$ alloy, relaxed or not.


%

\bibliographystyle{IEEEtran}
\bibliography{GeSn}

\section*{Appendix}
The band structure calculations were performed by using the Vienna $ab$ $initio$ simulation package (VASP) \cite{PhysRevB.54.11169} within the generalized gradient approximation (GGA) in Perdew-Burke-Ernzerhof (PBE) \cite{PhysRevLett.77.3865} type and the projector augmented-wave (PAW) pseudopotential \cite{PhysRevB.50.17953}. The kinetic energy cutoff is set to 560 eV, and the $k$-point grid was $12\times12\times12$ \cite{PhysRevB.13.5188}. The crystal structure is fully relaxed until the residual forces on atoms are less than 0.01 eV/\AA.

\section*{Acknowledgment}

 Weijun Fan acknowledges the funding support (NRF--CRP19--2017--01). The computation of this work was partially performed on resources
of the National Supercomputing Centre, Singapore. Greg Sun
acknowledges the grant support (FA9550-17-1-0354) from the Air Force Office of Scientific
Research.

\ifCLASSOPTIONcaptionsoff
  \newpage
\fi

\end{document}